\documentclass[9pt,twocolumn,twoside]{pnas-new-no-footer}

\usepackage{amssymb} 
\usepackage{mathtools} 
\usepackage{caption}
\usepackage{wrapfig}
\templatetype{pnasmathematics} 
\setboolean{displaywatermark}{false} 

\title{Detailed Balanced Chemical Reaction Networks as Generalized Boltzmann Machines}

\author[a, 1]{William Poole}
\author[b]{Tom Ouldridge} 
\author[c, 2]{Manoj Gopalkrishnan}
\author[a, 1, 2]{Erik Winfree}

\affil[a]{Computation and Neural Systems, California Institute of Technology, Pasadena, California, USA}
\affil[b]{Bioengineering, Imperial College London, London, England}
\affil[c]{Electrical Engineering, India Institute of Technology Bombay, Mumbai, India}

\leadauthor{Poole}

\significancestatement{This work presents a highly interdisciplinary approach towards understanding how biochemical systems can perform inference. By combining theories from the fields of machine learning, chemical reaction networks, and statistical physics, we have been able to provide a unique lens to understand biochemical networks as physical implementations of machine learning algorithms. In this work, we focus on inference: namely how can a chemical system make decisions based upon noisy data. Our results indicate that thermal fluctuations can be used as a resource at the micron scale. Finally, due to the concrete nature of our theory, we can directly apply known results from thermodynamics to chemical systems performing inferential computations.}

\authorcontributions{WP did the primary research, simulation, and wrote the manuscript. TO, MG, and EW jointly provided critical feedback, guidance, and technical expertise to the project.}
\authordeclaration{We declare no competing interests.}
\equalauthors{\textsuperscript{2} E.W. and M.G. jointly supervised this work.}
\correspondingauthor{\textsuperscript{1}To whom correspondence should be addressed. E-mail:
\texttt{wpoole@caltech.edu}; \texttt{winfree@caltech.edu}
}


\begin{abstract}
Can a micron sized sack of interacting molecules understand, and adapt to a constantly-fluctuating environment? Cellular life provides an existence proof in the affirmative, but the principles that allow for life's existence are far from being proven. One challenge in engineering and understanding biochemical computation is the intrinsic noise due to chemical fluctuations. In this paper, we draw insights from machine learning theory, chemical reaction network theory, and statistical physics to show that the broad and biologically relevant class of detailed balanced chemical reaction networks is capable of representing and conditioning complex distributions. These results illustrate how a biochemical computer can use intrinsic chemical noise to perform complex computations. Furthermore, we use our explicit physical model to derive thermodynamic costs of inference.
\end{abstract}

\dates{This manuscript was compiled on \today}

\begin{document}

\maketitle
\ifthenelse{\boolean{shortarticle}}{\ifthenelse{\boolean{singlecolumn}}{\abscontentformatted}{\abscontent}}{}

\dropcap{C}omputing with small numbers of molecules at around room temperature presents a unique set of challenges. However, cell and molecular biology demonstrate that every living cell can perform complex information processing using circuitry built out of just a few basic building blocks~\cite{alberts2015essential}. More recently, systems biologists have identified a variety of biochemical design principles observed across many organisms which suggests that unified understanding of biochemical systems may be possible~\cite{alon2019introduction,del2016have}. At the same time, synthetic biologists and engineers, inspired by the sophistication seen in biology, are beginning to build cellular and synthetic cell-like systems capable of functions ranging from locomotion to division to digital and analog information processing ~\cite{siuti2013synthetic,santiago2019self,aoki2019universal,xu2019protocell}. Programmed biochemical systems are rapidly expanding into medical diagnostics~\cite{slomovic2015synthetic}, cancer therapeutics~\cite{wu2019engineering}, sustainable bioreactors~\cite{french2019harnessing}, and advanced materials~\cite{le2018living}. However, designing and understanding increasingly complex biochemical computation will require new principles and design methodologies. This work adds a new perspective by showing that a wide class of biochemical models can be formally interpreted, designed, and analyzed using machine-learning inspired methods.

Almost all modern computers are built on the digital abstraction of variables having binary values. This abstraction allows computers to excel at tasks like Boolean logic and integer arithmetic. However, there is no reason why biochemical computers need be digital or would have evolved that way. Despite early success building synthetic biochemical logic circuits~\cite{qian2011scaling,nielsen2016genetic}, so far these systems have failed to achieve anything close to the complexity of a smartphone, much less a living organism. An alternative is to take an analog approach where biochemical signals are allowed a continuous range of values. In the past decades, analog computing has been reborn under the guise of deep learning and is revolutionizing domains from computer vision~\cite{krizhevsky2017imagenet} to natural language processing~\cite{brown2020language} to biochemistry~\cite{jumper2021highly}. A third approach, and the one we emphasize in this paper, is probabilistic programming~\cite{gordon2014probabilistic}, a hybrid of the conventional digital abstraction and analog computing where discrete values are assigned probabilities. Indeed, many state of the art machine learning methods involve probabilistic elements~\cite{ghahramani2015probabilistic} and more conventional machine learning methods may be seen as approximations of probabilistic models~\cite{ranganath2015deep}. It has also been suggested that analog or hybrid digital-analog approaches to information processing may be more efficient, particularly in the nervous system and biochemical contexts ~\cite{sarpeshkar1998analog,sarpeshkar2014analog}. There is also an increased interest in specialized hardware capable of directly implementing probabilistic computations and machine learning~\cite{james2017historical}. In this vein, machine learning architectures, such as neural networks, have been theorized~\cite{rossler1974synthetic,hjelmfelt1991chemical,mjolsness1991connectionist,buchler2003schemes,kim2004neural} and built in a variety of biochemical systems~\cite{cherry2018scaling} yet they remain considerably less complex than their digital relatives.

Inside individual cells, noise also plays an important role in biochemical computation which differs from the deterministic precision of silicon computers. At the micron to sub-micron scale, the counts of individual (non-solvent) molecules frequently become very small which in turn results in high variance~\cite{lestas2010fundamental}. Mathematically, such systems are commonly modeled with stochastic chemical reaction networks (CRNs)~\cite{gillespie1992rigorous}. Biochemical noise has been rigorously quantified in systems such as gene expression~\cite{elowitz2002stochastic}, stochastic partitioning at cell division~\cite{huh2011non}, and neuronal firing~\cite{faisal2008noise}. This extensive documentation of noise suggests that chemical computers must function in a a noisy environment, respond to noisy signals, and rely on noisy components. One way to deal with this noise is to mitigate it: methods such as kinetic proofreading~\cite{hopfield1974kinetic,evans2017physical}, low pass filters~\cite{hooshangi2005ultrasensitivity,fujita2010decoupling}, and fold change detectors~\cite{goentoro2009incoherent,kim2014synthetic} have all been observed naturally and engineered in order to mitigate fluctuations. However, an alternate approach is to ask how biochemical systems can use intrinsic noise to their advantage. A number of theoretical biochemical algorithms have been proposed that use intrinsic chemical fluctuations to generate distributions~\cite{cappelletti2020stochastic}, infer parameters of probabilistic models~\cite{virinchi2018reaction}, and solve combinatorial constraint satisfaction problems~\cite{winfree2019chemical}.

Probabilistic inference can be seen as the archetypal problem for understanding how to deal with noise and make the \textit{best} decision~\cite{shatcher1992Decision}. In computer science and machine learning, generative probabilistic models have emerged as a powerful framework for inference~\cite{jordan2004introduction}. Boltzmann machines are one of the most well studied of these models and excel at learning high dimensional probability distribution using hidden (latent) variables and computing conditional distributions~\cite{hinton1984boltzmann}. In this paper, we build on previous work which presented a number of CRN implementations of Boltzmann machines~\cite{poole2017chemical}. Here, we show how a broad class of stochastic biochemical models called detailed balanced (db) CRNs may be viewed as generalizations of Boltzmann machines: capable of representing complex distributions using hidden variables and computing conditional distributions. Due to the explicit and physical nature of these models, we then go on to provide some thermodynamic costs related to inference. In the a separate paper we apply these models to the task of learning. 

\section{Background}
\subsection{Probabilistic Inference}
In this work, we are interested in generative models, a broad class of mathematical models that represent probability distributions~\cite{salakhutdinov2015learning}. Specifically we focus on Boltzmann machines and stochastic chemical reaction networks; when interpreted as generative models, each of these systems probabilistically samples (or explores) a high dimensional state space. In general, we are less concerned with the sampling dynamics than the steady state distribution which occurs in the limit of running a generative model for infinite time. In fact, a low number of dynamic variables can represent a steady-state distribution that is exponential in the degrees of freedom~\cite{poole2017chemical}. Seen in this light, probability distributions present a powerful framework for a system to understand its environment. Within the context of generative models, we equate inference with the computation of conditional steady state distributions. For instance, a cell inferring the state of the environment $h$ given a noisy signal $v$ is equivalent to computing the conditional distribution $\mathbb{P}(h \mid v).$ Implicit in our framework is the idea that the variable being conditioned on is held constant while the conditional distribution is being computed - the constant variables are referred to as \textit{clamped}. Additionally, the un-clamped or \textit{free} variables must be coupled to the clamped variables for inference to be meaningful. In the limit of no coupling between $x$ and $y$, a distribution can be written in product form, $\mathbb{P}(x, y) = \mathbb{P}(x)\mathbb{P}(y)$, which means $x$ and $y$ are independent and conditioning provides no new information: $\mathbb{P}(x \mid y) = \mathbb{P}(x)$ and $\mathbb{P}(y \mid x) = \mathbb{P}(y)$.

Most readers will be familiar with conditional distributions of the form $\mathbb{P}(h \mid v)$ which is conditioning on $v$ having an exact value. However, it is also possible to generalize conditioning to holding moments of a distribution fixed via the conditional limit theorem \cite{cover1999elements}. For example, $\mathbb{P}(h, v \mid \langle v \rangle = \overline{v})$ means that the distribution $\mathbb{P}(h, v)$ has been conditioned on $v$ having an average value of $\langle v \rangle = \overline{v}$. This kind of conditioning allows $v$ to fluctuate, subject to constraints. Under a number of reasonable conditions we detail in Supplemental Section \ref{SI:proof_of_conditional_limit}, the conditional limit theorem states:
\begin{equation}
    \mathbb{P}(h, v \mid \langle v \rangle = \overline{v}) = \underset{P \in \mu}{\textrm{argmin }} \mathbb{D}( P(h, v) \mid \mid \mathbb{P}(h, v)).
\end{equation}
Here $\mathbb{D}(P \mid \mid Q) = \sum_x P(x) \log \frac{P(x)}{Q(x)}$ is the relative entropy and $\mu$ is the space of all distributions with mean $\langle v \rangle = \overline{v}$. As we will discuss, this generalization turns out to be more mathematically tractable than holding $v$ fixed and allows for complex inferential computations. Finally, in this work, we will use the term \textit{clamped} to also denote variables whose means have been conditioned and it will be clear by context which kind of conditioning is being performed.

\subsection{Boltzmann Machines}
\label{sec:BMs}
Boltzmann Machines\footnote{Many readers may be familiar with \textit{restricted} Boltzmann Machines which are a subclass of Boltzmann Machines with specific topological structure.} (BM) are a class of probabilistic graphical model which have been extensively studied theoretically~\cite{ackley1985learning,sejnowski1986higher} and used for many machine learning applications~\cite{salakhutdinov2009deep,tang2012robust,wang2013predicting}. For the purposes of this work, BMs serve as a guide for understanding what it means for a system to be capable of inference. Briefly, a BM is a stochastic neural network of binary nodes $x_i \in \{0, 1\}$ which have an equilibrium distribution:
\begin{equation}
    \mathbb{P}(x) = \frac{1}{Z} e^{-E(x)}  \quad \quad Z = \sum_{x \in \{0, 1\}^N} e^{- E(x)} \quad \quad E(x) = \sum_{i>j} w_{ij} x_i x_j - \sum_i \theta_i x_i.
\end{equation}
Here, $\theta \in \mathbb{R}$ are bias terms and the weight terms $w_{ij} \in \mathbb{R}$ couple nodes $x_i$ and $x_j$ and induce correlations ensuring $x_i$ and $x_j$ are not independent. Additionally Boltzmann machines can use hidden units and marginalization to represent even more complex distributions. Let $X = \{x_i\}$ be partitioned into two disjoint sets $V$ and $U$ of visible and hidden units, respectively. The probability of the visible units is given by the marginalization over the hidden units:
\begin{equation}
    \mathbb{P}(v) = \sum_u \mathbb{P}(v, u).
\end{equation}
Similar to the many layers of latent variables in a deep neural network, the hidden units $U$ have been shown to increase the complexity of the distributions that the BM can model~\cite{ackley1985learning,sejnowski1986higher,le2008representational}.

BMs are also able to seamlessly compute conditional distributions. Again, partition the nodes $X$ into two disjoint sets $U$ and $V$. Here $U$ will be the free variables and $V$ will be clamped. The conditional distribution $\pi(u \mid v)$ can be exactly sampled by clamping $v$ to a constant value while simulating the BM~\cite{ackley1985learning}. In many cases, a combination of clamping and marginalization is used together where both visible and hidden units can be clamped or free. General purpose inference allows Boltzmann Machines to classify data~\cite{larochelle2012learning} ($U$ are class labels and hidden units, $V$ are data points); generate distributions~\cite{taylor2009factored} ($U$ are data and hidden units, $V$ are labels and hidden units); and infer missing data~\cite{tang2012robust} ($U$ are unknown data and hidden units, $V$ are known data).

Finally, the parameters (energies) of a BM can learned via gradient descent on the relative entropy between a BM run freely and the distribution of the BM clamped to samples from data distribution $\psi(v)$.
\begin{align}
    \label{eq:learning_rule}
    &\frac{\textrm{d} w_{ij}}{\textrm{d} t} = \frac{\partial \mathbb{D}( \overline{\mathbb{P}} \mid \mid \mathbb{P})}{\partial w_{ij}} = \epsilon ( \langle x_i x_j \rangle_{\overline{\mathbb{P}}} - \langle x_i x_j \rangle_{\mathbb{P}}) 
    &&\frac{\textrm{d} \theta_i}{\textrm{d} t} = \frac{\partial \mathbb{D}( \overline{\mathbb{P}} \mid \mid \mathbb{P})}{\partial \theta_i} = \epsilon( \langle x_i \rangle_{\overline{\mathbb{P}}} - \langle x_i \rangle_{\mathbb{P}}) 
\end{align}
Here, $\overline{\mathbb{P}}(v, u) = \mathbb{P}(u \mid v) \psi(v)$ is the clamped distribution and $\epsilon$ is the learning rate. Importantly, this optimization procedure is equivalent to minimizing the relative entropy $D\mathbb{D}(\psi(v) \mid \mid \mathbb{P}(v))$. However, by using the clamped distribution $\overline{\mathbb{P}}$ this learning rule is able to train hidden units representing latent variables not part of the data distribution $\psi$. 

\subsection{Chemical Reaction Networks}
In this work, we model biochemical fluctuations using stochastic chemical reaction networks (CRNs) with mass action kinetics. This model is consistent with the statistical mechanics of a well-mixed ideal solution~\cite{gillespie1992rigorous} and has a long history being applied to biological problems including modeling genetic circuits~\cite{murray2010biomolecular} and understanding noise in biochemical processes~\cite{lestas2008noise}. Similarly, stochastic CRNs have been studied as a programming language, shown to be Turing universal~\cite{soloveichik2008computation,cook2009programmability,cummings2016probability} and used to guide implementations and analyses of molecular programs in laboratory settings~\cite{chen2013programmable,srinivas2017enzyme,badelt2017general} .

Formally, a CRN $(\mathcal{S}, \mathcal{R}, k)$ is a set of species $\mathcal{S}$, reactions $\mathcal{R}$ and reaction rates $k$. Reactions convert one multiset of species into another: $I^r \xrightarrow{k_r} O^r$. Here $I^r$ and $O^r$ are the vectors of input and output species for reaction $r$, respectively. These vectors can be combined into a single matrix called the stoichiometric matrix $M = O - I$. CRNs may have both stochastic and deterministic dynamics. However, in this paper we focus on the stochastic dynamics of the probability that species in the CRN will have particular counts. Reactions occur with probability proportional to their mass-action propensity function $\rho_r(s) = \frac{k_r}{\mathbb{V}^{|I^r|}} \prod_i \frac{s_i !}{(s_i-I_i^r)!}$. Here, $\mathbb{V}$ denotes the volume of the CRN and will be assumed to be 1 unless explicitly defined otherwise and $k_r$ have units of per second. We use $s_i$ to mean the counts of species $S_i$ and $s$ without a subscript denotes a vector of species' counts. The dynamics of the probability distribution are given by the chemical master equation:
\begin{equation}
    \label{eq:cme}
    \frac{\textrm{d}\mathbb{P}(s, t)}{\textrm{d}t} = \sum_r \mathbb{P}(s-M^r, t)\rho_r(s-M^r) - \mathbb{P}(s, t)\rho_r(s).
\end{equation}
Often, we are interested in the stationary distribution $\mathbb{P}^*(s; s^0)$ found by setting (\ref{eq:cme}) to 0 or by simulating the CRN dynamics until convergence with exact methods such as the Gillespie algorithm~\cite{gillespie2007stochastic}. For arbitrary CRNs, $\mathbb{P}^*(s; s^0)$ may or may not exist, may not be unique, and may depend on the initial condition $s^0$. However, for the class of detailed balanced CRNs defined below existence and uniqueness are guaranteed.

The reachability class of a CRN, $\Gamma_{s^0} \subseteq \mathbb{Z}^{\mathcal{S}}$, is the subset of the integer lattice reachable by a sequence of reactions starting at an initial state $s^0$. In some cases, $\Gamma_{s^0}$ may be infinite. We emphasize that this set is distinct from the stoichiometric subspace $\Omega_{s^0}$ which is given by the kernel of the stoichiometric matrix $M$. The latter is an affine space of the form $A (s - s^0) = 0$, with $A$ a $|\mathcal{S}|$ by $k$ dimensional matrix representing $k \geq 0$ conserved quantities in the system. Specifically, $\Gamma_{s^0} \subseteq \Omega_{s^0}$~\cite{gunawardena2003chemical}. For example, the CRN $\emptyset \xrightleftharpoons[]{} 2 S$ has an infinite reachability class of either the even or odd positive integers depending on $s^0$ and a stoichiometric compatibility class that covers \textit{all} the positive integers. Indeed, as we will prove later, it is by restricting the reachability class relative to the stoichiometric subspace that we are able to program detailed balanced CRNs to represent broad classes of distributions. Note that the reachability class may be arbitrarily complex compared to the stoichiometric compatibility class~\cite{leroux2019reachability}.

\subsection{Detailed Balanced Chemical Reaction Networks}
Detailed balanced CRNs (dbCRN) are a subclass of CRNs that represent non-driven chemical systems such as molecular binding and unbinding. Mathematically, the detailed balanced property requires that:
\begin{itemize}
    \item Each species $S_i \in \mathcal{S}$ has an energy $G_i$
    \item All reactions are reversible meaning if $I \xrightarrow{k^+} O \in \mathcal{R}$ then $O \xrightarrow{k^-} I \in \mathcal{R}$.
    \item Reaction rates obey $\frac{k^+}{k^-} = e^{-\Delta G}$, where $\Delta G = \sum_i G_i(O_i- I_i)$.
\end{itemize}
A detailed balanced reaction network's stationary distribution is an equilibrium distribution meaning there is no net energy flow or entropy production at steady state. These distributions can correspond to both the canonical or grand canonical ensembles of statistical physics, depending on whether only energy is allowed to be exchanged with the surrounding heat bath or if certain species within a reservoir can also be exchanged (such as via the reaction  $\emptyset \xrightleftharpoons[]{} 2 S$ where $\emptyset$ indicates a molecule moving to or from the reservoir). This equilibrium  distribution necessarily has product a Poisson form written in terms of the free energy function $\mathcal{G}(s)$~\cite{anderson2010product}:
\begin{equation}
    \label{eq:product_poisson}
    \pi(s; s^0) = \frac{1}{Z}\prod_i \frac{e^{-G_is_i}}{s_i!} = \frac{1}{Z} e^{-\mathcal{G}(s)} \quad \quad Z = \sum_{s \in \Gamma_{s^0}} e^{-\mathcal{G}(s)} \quad \quad \mathcal{G}(s) = \sum_i \mathcal{G}_i(s) =  \sum_i G_i s_i + \log s_i !
\end{equation}
Notice that in general the equilibrium distribution $\pi(s; s^0)$ depends on the initial condition $s^0$. However, for notational simplicity we will typically just write $\pi(s)$ with the initial condition dependence implied. Similarly, the sum in the partition function runs over the reachability class $\Gamma_{s^0}$ and the dependence of $Z$ on $s^0$ is always implied. Additionally, in some proofs to avoid cluttered notation, the reachability class may be omitted from sums over species $s$ but is always implied. We comment that $\pi(s; s^0)$ is ergodic within the reachability class $\Gamma_{s^0}$ due to the detailed balanced conditions. Finally, in this paper, we are primarily concerned with how species' energies affect the equilibrium distribution. If $\mathcal{D} = (\mathcal{S}, \mathcal{R}, k)$ is a detailed balanced CRN, $\mathcal{D}_G$ is the same set of species and reactions with the rate constants $k$ changed to reflect the new energies $G$. This CRN will not necessarily be unique because any pair of forward and backward rates can be rescaled arbitrarily, however it will have a unique equilibrium distribution given by equation (\ref{eq:product_poisson}).

\subsection{Detailed Balanced CRNs Can Model Complex Environments}
\label{sec:dbcrns_model_complex_environments}
\begin{figure}
    \centering
    \includegraphics[width=.9\textwidth]{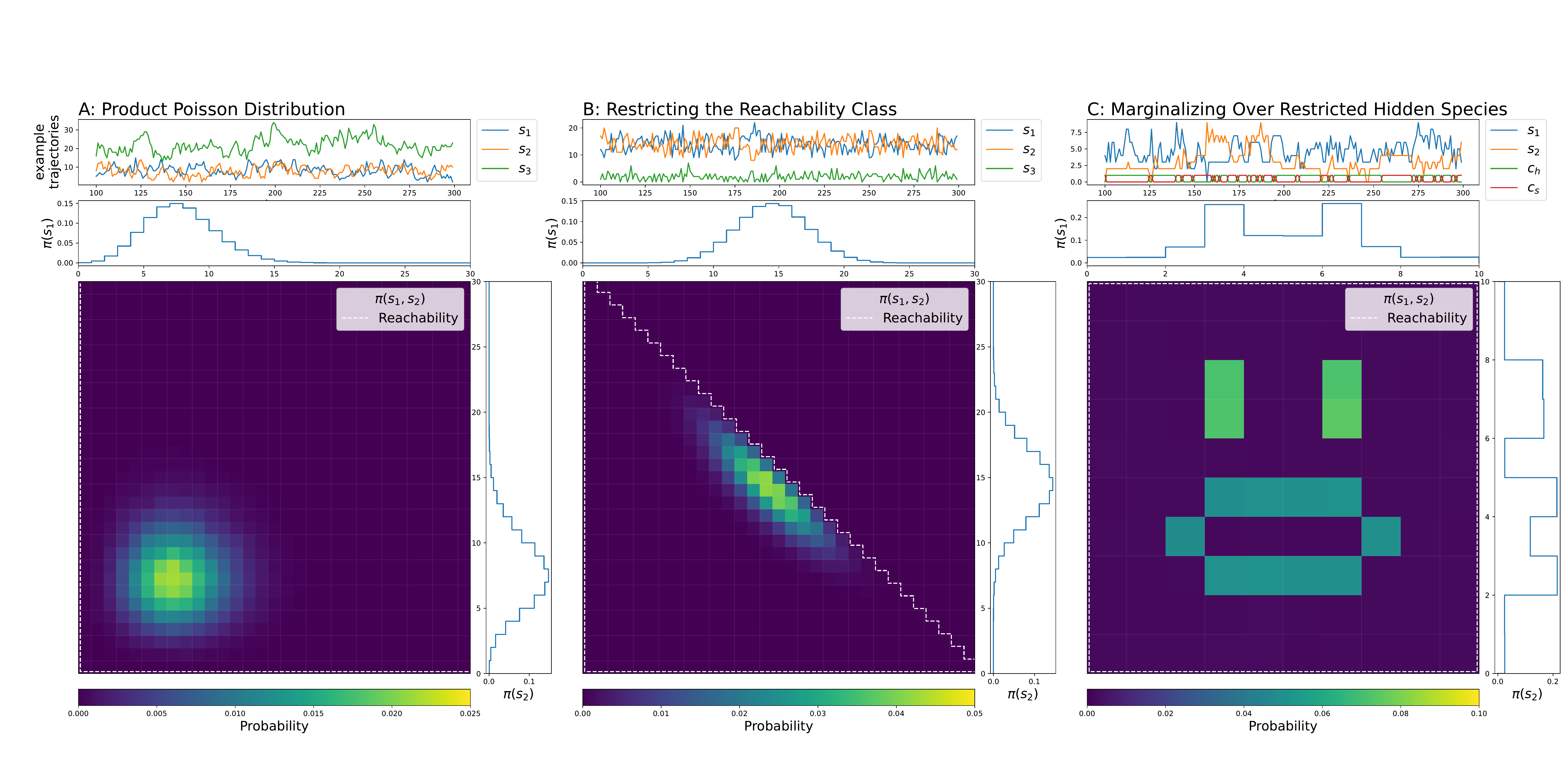}
    \caption{Restricting the reachability class of a dbCRN can give rise to complex distributions. A. Product Poisson Distribution of two independent species marginalized over a third independent species generated from a dbCRN whose reachability class is the entire positive integer lattice: $\emptyset \xrightleftharpoons[]{} S_i$. B. The marginal equilibrium distribution of a dbCRN restricted whose reachability class is restricted to an affine subspace: $S_1 \xrightleftharpoons[]{} S_2 \xrightleftharpoons[]{} S_3$. This reachability class induces correlations between the species. C. The marginal equilibrium distribution of a dbCRN with a highly restricted reachability class marginalized over many binary hidden species. See equation (\ref{eq:pixel_crn}) for the full dbCRN. This dbCRN produces a distribution with a very complex structure. Note: all trajectories shown at the top are just tiny fractions of the total simulated time.}
    \label{fig:dist}
\end{figure}

Equation (\ref{eq:product_poisson}) seems to suggest that dbCRNs can only produce simple distributions in product-Poisson form that seem to only allow control of the mean of the species' counts, because Poisson distributions are defined by their mean. For example, a product Poisson distribution on the entire integer lattice can be seen in figure \ref{fig:dist}A. Yet, in previous work, we constructed dbCRNs able to produce non-Poisson distributions such as those produced by Boltzmann Machines~\cite{poole2017chemical}. Indeed, carefully constructed dbCRNs can in fact produce any distribution with finite support~\cite{cappelletti2020stochastic}. Each of these constructions implicitly restricts the reachability class with specific reactions and initial conditions. In this section, we  illustrate two example of dbCRNs producing non-poisson distributions. First, we consider the simple dbCRN used to produce the distributions in Figure \ref{fig:dist}B:
\begin{align}
    &S_1 \xrightleftharpoons[]{} S_2 \xrightleftharpoons{}{} S_3 &&s_1 + s_2 + s_3 = c.
\end{align}
Here the a single constraint restricts the reachability class of each species where $c$ is the total number of all the species at the start of the simulation. Even such a simple constraint is enough to break the independence of a product Poisson distribution which can be seen by the obvious anti-correlatiom between $S_1$ and $S_2$ in Figure \ref{fig:dist}B.

Detailed balanced CRNs with many more conservation laws can produce arbitrary distributions with finite support. For example, the dbCRN used to produce the face in figure \ref{fig:dist}C and the smiley-frowny face in \ref{fig:energy_clamping}C is given by the reactions:
\begin{align}
\label{eq:pixel_crn}
     C_\alpha + P^\alpha_{x, y} \xrightleftharpoons[]{} C_\alpha+P^\alpha_{x+1, y} + S_1,
     \quad \quad
     C_\alpha + P^\alpha_{x, y} \xrightleftharpoons[]{} C_\alpha P^\alpha_{x, y+1} +S_2,
     \quad \quad
     C_\alpha + P^\alpha_{x, y} \xrightleftharpoons[]{} C_\beta + P^\beta_{x, y}.
\end{align}
Here, $C_\alpha$ are control species tuning whether the distribution is happy  ($\alpha = h$) or sad ($\alpha = s$). There are two sets of pixel species $P_{x, y}^\alpha$, one for the happy face ($\alpha = h$) and one for the sad face ($\alpha = s$) with $x$ and $y$ denoting the pixel locations. The visible species are $S_1$ and $S_2$. The energies of the pixel species have been set to produce the happy and sad images. This construction requires a very tightly coupled reachability class with initial conditions that satisfy the following constraints: 
\begin{align}
    c_s + c_h = 1 \quad \quad \sum_{x, y, \alpha} p_{x, y}^\alpha = 1, \quad \quad 
    \sum_{x, y} p_{x, y}^\alpha = c_\alpha,
    \quad \quad 
    s_1 = \sum_{x, y, \alpha} x p_{x, y}^\alpha \quad \quad s_2 = \sum_{x, y, \alpha} y p_{x, y}^\alpha.
\end{align}
In words: there is only one $C_\alpha$ species present and one $P_{x, y}^\alpha$ species present at any time and they must both have the same value for $\alpha$. The counts of $S_1$ and $S_2$ must also correspond to the values $x$ and $y$, respectively, of the single $P_{x, y}^\alpha$ species present. Importantly, if this dbCRN begins in a state which obeys these conservation laws, all subsequent states will also obey these laws due to the carefully chosen set of reactions.

\subsection{Chemical Boltzmann Machines}
\label{sec:cbms}
In our previous work on chemical Boltzmann machines~\cite{poole2017chemical}, we showed that a special class of detailed balanced CRNs called the \textit{Edge-Species Chemical Boltzmann Machine} (ECBM) construction exactly simulates a traditional Boltmzann Machine and is capable of a limited form of inference. This construction uses reactions to constrain the reachability class of certain species with conservation laws. This fact can be seen by the example of a two-node ECBM (which can be generalized to arbitrary graphs):
\begin{align}
    S_1^0 + S_2^0 \xrightleftharpoons[]{} S_1^1 + S_2^0 \quad \quad S_1^1 + S_2^0 + S_W^0 \xrightleftharpoons[]{} S_1^1 + S_2^1 + S_W^1 \\
    S_1^0 + S_2^0 \xrightleftharpoons[]{} S_1^0 + S_2^1 \quad \quad S_1^0 + S_2^1 + S_W^0 \xrightleftharpoons[]{} S_1^1 + S_2^1 + S_W^1.
\end{align}
Here, each node $i \in \{1, 2\}$ of a BM is represented by two species $S_i^0$ and $S_i^1$ corresponding to the \textit{off} and \textit{on} states. The edge species $S_W^0$ and $S_W^1$ relates to the energy term $w_{12}$ in a BM and similarly has \textit{off} and \textit{on} states. This dbCRN has stoichiometric conservation laws $s_i^0 + s_i^1 = N_i \quad i \in \{1, 2, W\}$ as well as the possibility of an emergent conservation law $s_0^1 s_1^1 = s_W^1$ which occurs when $N_1=N_2=N_W=1$ and the CRN starts in a state which respects this law. The key observation is that, due to the this final conservation law, the BM energy $w_{ij}$ directly corresponds to the species energy $G_{W_{ij}^1}$ of an ECBM. Additionally, in our past work, we developed a rudimentary clamping process by which a subset of species in the dbCRN is held constant for conditioning (possibly "turning off" a subset of reactions when some species are clamped to zero). However we observe that our previous clamping construction is fragile because it requires that the reachability class of the unclamped species remains unchanged, which will only be true for very carefully designed dbCRNs.

In this work, we generalize the ideas of clamping to all dbCRNs and show how they can be implemented in a purely chemical setting. First, we will provide a general framework for understanding how restricting the reachability class of a dbCRN enables the production of complex distributions. Then, we will provide a new definition of clamping which is broadly applicable to any dbCRN demonstrating that any dbCRN is capable of inference. Finally, we will use this notion of clamping to derive some thermodynamic costs for inference.
\section{Effective Use of Hidden Species Requires Reachability Entanglement}
In this section, we provide the first core result of this paper: a unifying framework for understanding how some dbCRNs are able to produce equilibrium distributions which are far from products of independent Poisson distributions. First, we will show that restricting the reachability class of dbCRNs is essential to producing far-from-Poisson distributions. Furthermore, when using hidden units to increase the representational power of a dbCRN, the reachability class of these units must be ``entangled'' with the reachability class of the visible units for the hidden units to have any effect on the visible distribution.

As a reference case, consider a dbCRN with reactions such that the entire positive integer lattice is reachable: $\Gamma^\rho_{s^0} = \mathbb{Z}_{\geq 0}^{|S|}$ with equilibrium distribution $\rho(s)$. Then all the species are independent of each other because $\rho$ is a product of independent Poisson distributions.
\textit{Proof:}
Rewrite equation (\ref{eq:product_poisson}) into product form to show independence.
\begin{align}
    \rho(s) &= 
    \frac{\prod_i e^{-\mathcal{G}_i(s_i)}}{\sum_{s \in \mathbb{Z}_{\geq 0}^n} \prod_i e^{-\mathcal{G}_i(s_i)}}
    = \frac{\prod_i e^{-\mathcal{G}_i(s_i)}}{\prod_i \sum_{s_i \in \mathbb{Z}_{\geq 0}}  e^{-\mathcal{G}_i(s_i)}}
    = \prod_i \frac{e^{-\mathcal{G}_i(s_i)}}{\sum_{s_i \in \mathbb{Z}_{\geq 0}} e^{-\mathcal{G}_i(s_i)}} = \prod_i \rho_i(s_i).
\end{align}
Notice that switching the order of the sum and product in the denominator is only possible because the entire positive integer lattice is reachable. An example of such a CRN, $\emptyset \xrightleftharpoons[]{} S_i \; i \in \{1, 2, 3\}$, is illustrated in figure \ref{fig:dist}A. In general, the partition function $Z$ cannot necessarily be factored this way. 

Now, consider a second dbCRN with the same species and the same species' energies but different reactions and a different reachability class $\Gamma^\pi_{s^0} \subset \Gamma^\rho_{s^0}$ and equilibrium distribution $\pi(s; s^0)$. How well a product poisson $\rho$ is approximated by the distribution $\pi$ can be measured using the relative entropy:
\begin{equation}
\label{eq:relative_entropy_scaling}
    \mathbb{D}(\pi \mid\mid \rho)
    = \sum_{s\in \Gamma_{s^0}^\rho} \pi(s) \log \frac{\pi(s)}{\rho(s)}
    = \sum_{s\in \Gamma_{s^0}^\rho} \frac{e^{-\mathcal{G}(s)}}{Z^\pi} \log \frac{Z^\rho}{Z^\pi} 
    = \log (\frac{Z^\rho}{Z^\pi})
    = \log (1 + \frac{Z^-}{Z^\pi}).
\end{equation}
Here, $Z^\rho$ and $Z^\pi$ are the partition functions of $\rho$ and $\pi$, respectively, and $Z^- = Z^\rho - Z^\pi>0$. Note that we are using the convention $\pi(s) = 0 \, \forall \, s \not \in \Gamma^\pi_{s^0}$ and $0 \log 0 = 0$. Due to the logarithm, for $\pi$ to be far from product Poisson, a heavily weighted subset of states must be unreachable so that $Z^- \gg  Z^\pi$. The figure \ref{fig:dist}B illustrates how restricting the reachability class with linear conservation laws induces correlations using the dbCRN is $S_1 \xrightleftharpoons[]{} S_2 \xrightleftharpoons[]{} S_3$ as an example.

More generally, the combination of marginalization and restricting the reachability class can be employed to increase the relative entropy between the equilibrium distribution of any dbCRN, $\sigma(v)$, and the equilibrium distribution of another dbCRN with the same species $V$ as well as additional hidden species $U$ such that the equilibrium is $\pi(v, u)$. We will show this by providing a sufficient condition for hidden units to \textit{not} increase the relative entropy and then describe how to construct systems which do not satisfy that condition. Assume that all the species $V$ have the same energies $G$ in both dbCRNs and that the reachable states in the first dbCRN are also reachable in the second dbCRN, for some value of the hidden species $u$: $\Gamma^\sigma_{v(0)} \cap \Gamma^\pi_{v(0), u(0)} = \Gamma^\sigma_{v(0)}$ where the initial condition is $s^0 = v^0, u^0$ (here $\cap$ is used loosely to denote the overlap between two spaces of different dimensions as opposed to a typical set intersection). The marginal distribution of the second dbCRN is:
\begin{equation}
    \pi(v) = \sum_{u \in \gamma(v)} \pi(v, u) \quad \quad \gamma(v) = \{u \textrm{ s.t. } v'=v \; \mid \; (u, v') \, \in \Gamma_{s^0}^\pi\}
\end{equation}
Here the function $\gamma$ produces a set of the reachable hidden states $u$ given a visible state $v$ with the dependence on the reachability class and initial condition implied in the function arguments for brevity. The relative entropy can then be written:
\begin{align}
    \mathbb{D}(\pi(v) \mid \mid \sigma(v)) 
    &= \sum_{v \in \Gamma^\sigma} \pi(v) \log \frac{\pi(v)}{\sigma(v)}
    = \sum_{v \in \Gamma^\sigma} \left ( \sum_{u \in \gamma(v)} \pi(v, u) \right ) \log \frac{\sum_{u \in \gamma(v)} \pi(v, u)}{\sigma(v)}\\
    &= \sum_{(v, u) \in \Gamma^\pi} \pi(v, u) \log \left (\frac{Z^\sigma}{Z^\pi} \sum_{u \in \gamma(v)} e^{-\mathcal{G}(u)}\right )
\end{align}
If $\gamma(v)$ is constant for all $v$, the restrictions on the reachability class are independent of the visible species, allowing the partition function to be factored $Z^\pi = Z^\sigma (\sum_u e^{-\mathcal{G}(u)})$. This simplifies the previous equation to 0 showing that marginalization requires restricted reachability classes to modify distributions:
\begin{align}
    \gamma(v) = \textrm{const} \implies \mathbb{D}(\pi(v) \mid \mid \rho(v)) 
    &= \log \frac{Z^\sigma ( \sum_{u} e^{-\mathcal{G}(u)}) }{Z^\pi} = 0.
\end{align}
A conceptual way to interpret this result is that marginalizing over the species $U$ does not affect the species $V$ if the reachable states of $V$ and $U$ are independent. This illustrates that complex reachability restrictions on the hidden species are not enough: for them to provide extra power for representing complex distributions, the reachability of the visible and hidden species must be entangled. Figure \ref{fig:dist}C depicts the equilibrium distribution of a dbCRN which marginalizes over many hidden species and extensively restricts reachability. This construction relied on very specific initial conditions described in section \ref{sec:dbcrns_model_complex_environments} which cause the hidden pixel species are highly entangled with each other---only one pixel can be present at a time---and also with the visible species---each visible count $(s_1, s_2)$ corresponds to a unique pixel species $P_{s_1, s_2}$.  

In summary, dbCRNs can represent a diverse set of distributions. When a dbCRN is unconstrained with species' counts allowed to take any value, all dbCRNs will have Poisson equilibrium distributions with species' means determined by their energies (illustrated in Figure \ref{fig:dist}A). Species may then be coupled together via conservation laws which can induce correlations (illustrated in Figure \ref{fig:dist}B). Furthermore, increasingly complex conservation laws have the potential to dramatically constrain the reachability class and enable the production of distributions with very rich structure. In some cases, the reachability class can be constrained via auxiliary hidden species which have their reachable states entangled with the reachable states of the visible species. In such cases, marginalization over the hidden species may produce even more complex distributions on the visible species (illustrated in Figure \ref{fig:dist}C). Finally, we comment that marginalization occurs implicitly when two chemical systems interact. Consider subsystem $A$ with species $\mathcal{S}^A = (V, U^A)$ which observes a different subsystem $B$ with species $\mathcal{S}^B = (V, U^B)$. If $A$ observes the species $V$ long enough, it implicitly observes the marginal distribution over the unobserved species $U_B$. 

\section{Inference with Detailed Balanced CRNs}
As referenced in the background section \ref{sec:BMs} on Boltzmann Machines, general purpose inference is incredibly powerful and can be used for a wide range of computational tasks. In the following sections, we show that dbCRNs are similarly capable of inference. To do this, we define a new notion of clamping which overcomes the limitations of our previous work~\cite{poole2017chemical} and is applicable to all dbCRNs. We then show how this clamping can be implemented by an auxilary set of chemical species. In this section, each CRN will be assumed to have its species partitioned into disjoint sets of free and clamped species, $\mathcal{S} = (\mathcal{S}^F, \mathcal{S}^C)$.

\begin{figure}
    \centering
    \includegraphics[width=.75\textwidth]{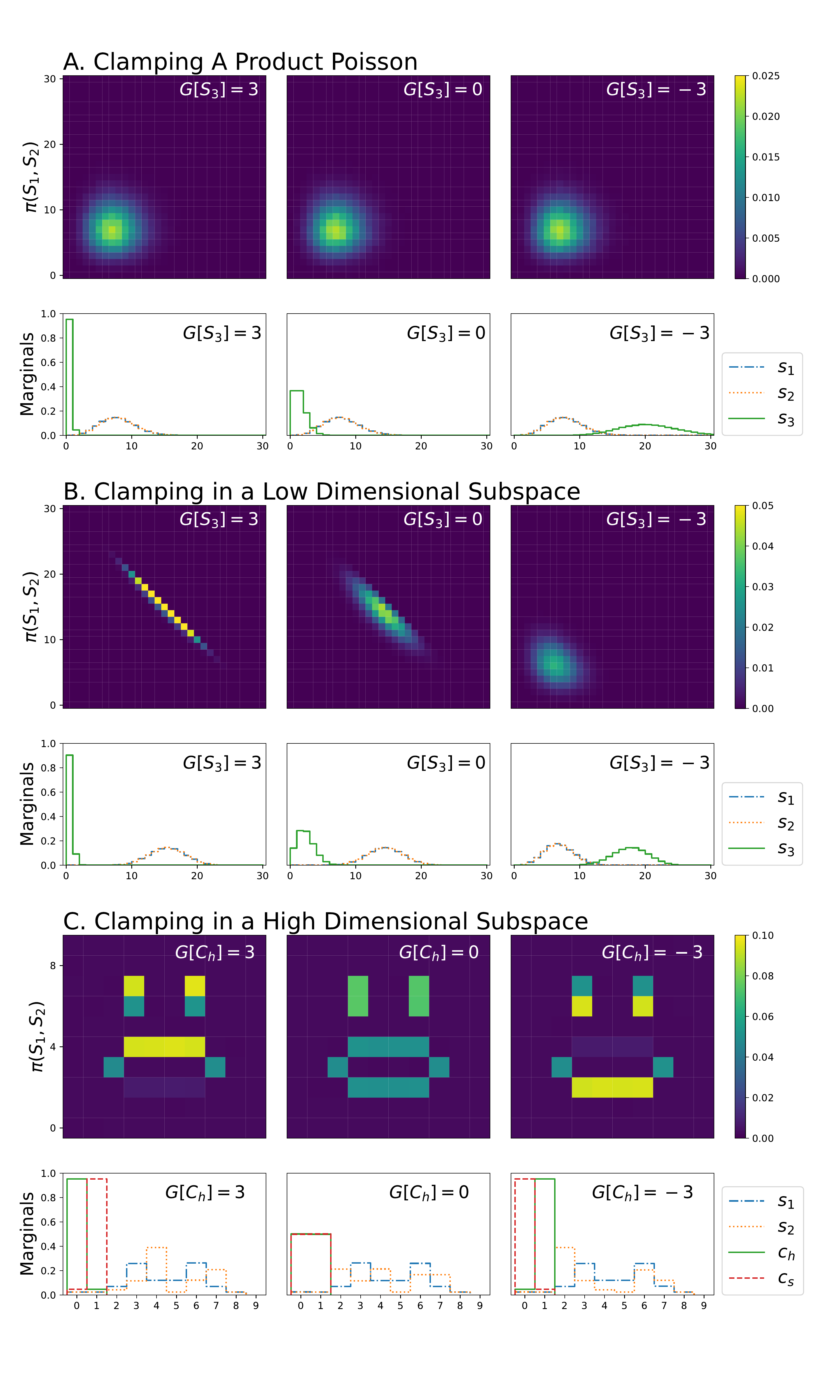}
    \vspace*{-15mm}
    \caption{Energy clamping hidden species is a powerful inferential computation capable of dramatically changing the marginal distribution of visible species. A. Energy clamping a dbCRN with a reachability class covering the entire positive integer lattice: $\emptyset \xrightleftharpoons[]{} S_i$. Clamping the energy of $S_3$ does not change the marginal distribution of $\pi(s_1, s_2)$. B. Energy clamping species in a dbCRN with a restricted reachability class can increase or decrease correlations in unclamped species. Here, the energy of $S_3$ in the dbCRN $S_1 \xrightleftharpoons[]{} S_2 \xrightleftharpoons[]{} S_3$ is clamped causing $S_1$ and $S_2$ become increasingly anti-correlated at high energies and more independent at low energies. C. Energy clamping species in a dbCRN with many hidden species and highly entangled reachability classes between hidden and visible species can give rise to dramatic changes in the marginal distribution of the visible species. Here, the clamping a single hidden species $C_h$ is capable of generating diverse distributions illustrated by the frowning, open mouthed, and smiling emojis. The construction for this dbCRN is described in by equation (\ref{eq:pixel_crn}).}
    \label{fig:energy_clamping}
\end{figure}

\subsection{Energy Clamping}

We define a new kind of clamping which is applicable to all dbCRNs. We call this process energy clamping because it works by modulating the energies of chemical species illustrated in figure \ref{fig:clamping_cartoon}A. The energy clamped dbCRN $\mathcal{D}_{G^C}$ has the same species and reactions as $\mathcal{D}_{G}$ but different energies and hence different reaction rates:
\begin{align}
    G^C_i = 
    \begin{cases}
    G_i \quad &S_i \in \mathcal{S}^F\\
    G_i + \Delta_i \quad &S_i \in \mathcal{S}^C
    \end{cases}
\end{align}
where $\Delta_i$ are the changes in energy for each clamped species. The clamped dbCRN has a new equilibrium distribution $\pi_C(s)$:
\begin{align}
    \label{eq:energy_clamp_equilibrium}
   \pi_C(s) = \frac{e^{-\mathcal{G}^C(s)}}{Z^C} \quad  \quad Z^C = \sum_{s \in \Gamma_{s^0}} e^{-\mathcal{G}^C(s)} \quad \quad
   \mathcal{G}^C(s) = \sum_i G^C_i s_i + \log s_i!
\end{align}
Here we are treating the energy changes $\Delta$ as tunable parameters which control the means of our species. Unlike traditional clamping of a Boltzmann Machine or the construction from the CBM paper, energy clamping does not hold the value of a species fixed. Instead, energy clamping can be interpreted as holding the mean of the clamped species fixed while allowing for fluctuations. These fluctuations are important because they ensure that the reachability class is preserved. This is formalized via the second core result of this paper which equates energy clamping to inference by conditioning on the mean via the conditional limit theorem~\cite{cover1999elements}: 
\\\\%
\textbf{Theorem:} Energy clamping is equivalent to conditioning upon the mean $\langle s^C \rangle$ being equal to $\overline{c}$.
\begin{equation}
    \label{eq:energy_clamping_conditional_limit}
    \pi_C(s^F, s^C) = \pi(s^F, s^C \mid  \langle s^C \rangle = \overline{c})
\end{equation}
Here, we have implicitly chosen $\Delta$ such that $\langle s^C \rangle_{\pi_C} = \overline{c}$. \textit{Proof:} See SI \ref{SI:proof_of_conditional_limit}. We note that there is generally no analytic formula for choosing the values of $\Delta$ which result in a particular mean; however, optimization procedures such as the gradient descent learning rule for dbCRN energies described in \cite{poole2017chemical} can be used to find these parameters numerically.
\\\\
The idea of conditioning on the mean versus holding the value of a species constant is analogous to the difference between microcanonical ensembles, where energy is constant, and the canonical ensemble where the mean energy is fixed, but allowed to fluctuate. Energy clamping modulates the energy of a species which holds the steady-state mean value constant but the actual count of that species is still allowed to fluctuate. The correctness of energy clamping can also be seen more simply through the following theorem on the conditional distributions:
\\\\%
\textbf{Theorem:} Energy clamping produces the same conditional distributions between $\pi$ and $\pi_C$ when conditioned on the species $S^C$ taking a specific value $c$:
\begin{equation}
    \pi(s^F \mid s^C = c) = \pi_C(s^F \mid s^C = c).
\end{equation}
\textit{Proof:}
\begin{align}
    \pi(s^F \mid s^C = c) 
    &= \frac{\pi(s^F, c)}{\pi(c)} 
    = \frac{e^{-\mathcal{G}(s^F) - \mathcal{G}(c)}}
    {\sum_{s^F} e^{-\mathcal{G}(s^F) - \mathcal{G}(c)}} = \frac{e^{-\mathcal{G}(s^F)}}
    {\sum_{s^F} e^{-\mathcal{G}(s^F)}} 
\end{align}

\begin{align}
    \pi_C(s^F \mid s^C = c) 
    &= \frac{\pi_C(s^F, c)}{\pi_C(c)} 
    = \frac{e^{-\mathcal{G}^C(s^F) - \mathcal{G}^C(c)}}
    {\sum_{s^F} e^{-\mathcal{G}^C(s^F) - \mathcal{G}^C(c)}} = \frac{e^{-\mathcal{G}(s^F)}}
    {\sum_{s^F} e^{-\mathcal{G}(s^F)}} 
\end{align}
here the final steps notes that $ \mathcal{G}^C(s^F) = \mathcal{G}(s^F)$ by definition. This result shows that the clamped dbCRN has the correct conditional distribution when conditioned on any $s^C = c$. 

Energy clamping provides a framework to hold the species of an arbitrary detailed balanced CRN around a value by modulating the energy vector $G \Rightarrow G^C$. In the case where $S_i$ can take any value on the integer lattice, energy clamping can be thought of as tuning the mean of $S_i$ directly as seen in the lower plot of figure \ref{fig:energy_clamping}A. In the case where $S_i$ is constrained via a conservation class to some minimum and maximum value, energy clamping may be better thought of as pushing $S_i$ towards one of its extreme values. Furthermore, when reachable states of the clamped species are entangled with the reachable states of free species, energy clamping induces changes in the distribution of free species illustrated clearly in figure \ref{fig:energy_clamping}B. In these cases, energy clamping may perform computationally challenging inferential tasks. For example, in Figure \ref{fig:energy_clamping}C, clamping a single species $C_h$ is able to induce dramatic changes in the distribution of the visible species causing the transition between a frowning face to a smiling face.  Importantly, the energy clamping construction does not change the reachability class of the underlying dbCRN which allows it to apply to any species in any dbCRN, regardless of reactions in the CRN. In the next section, we will describe how energy clamping can be implemented via external chemical potentials.

\subsection{Clamping with A Potential Bath}

\begin{wrapfigure}{R}{.4\textwidth}
    \centering
    \includegraphics[width=.39\textwidth]{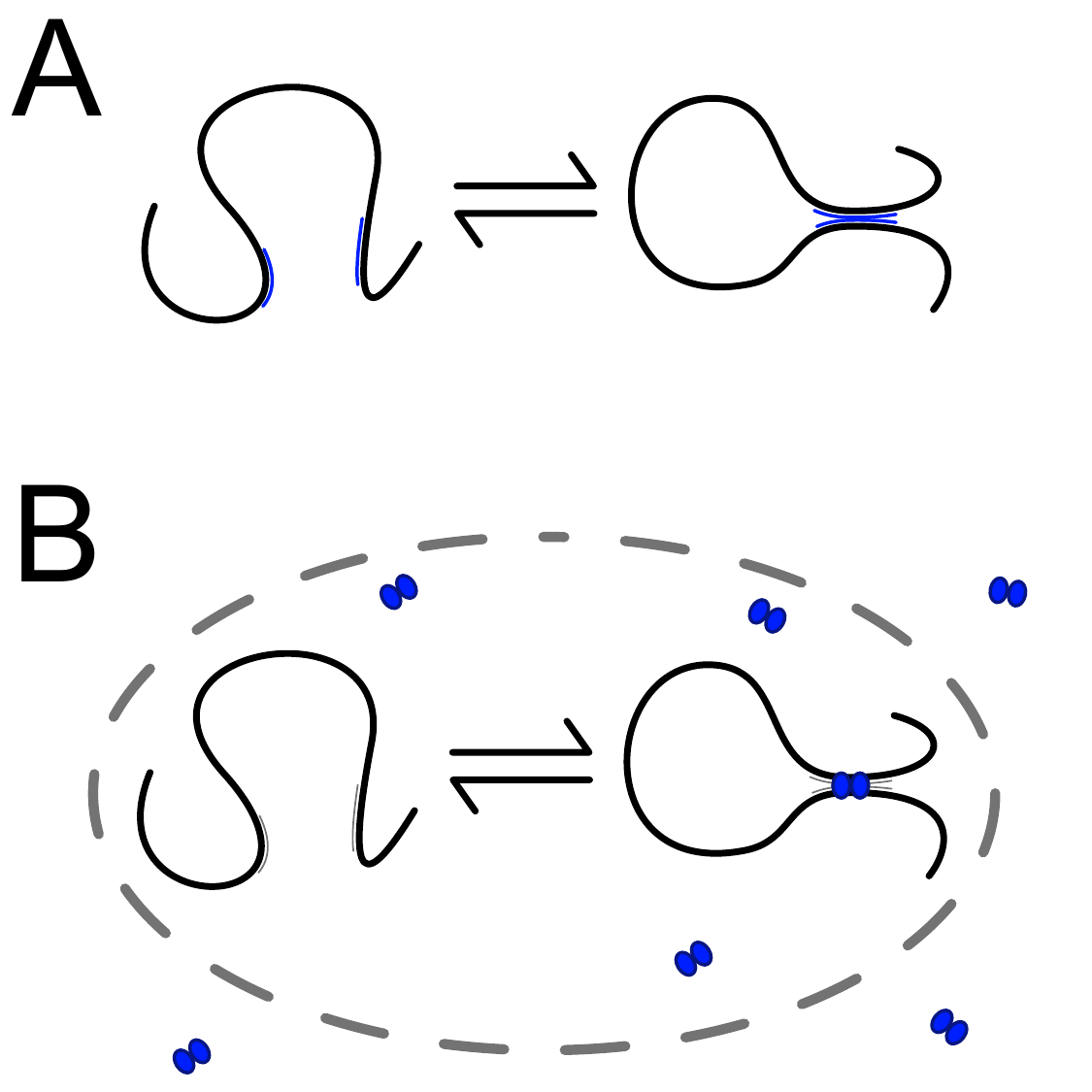}
    \caption{A. An illustration of energy clamping the reaction $S_{open} \xrightleftharpoons[]{} S_{loop}$. The reaction can be pushed in either direction by by changing the sequence depicted in blue which can tune the relative energies $\Delta_{S_{open}}$ and $\Delta_{S_{loop}}$. B. An illustration of potential clamping where $S_{open}$ is connected to a potential species $P$ resulting in the reaction: $P + S_{open} \xrightleftharpoons[]{} S_{loop}$. Changing $[P]$ is the equivalent to changing $\Delta$. The gray lines indicate that $S$ is in a small volume relative to $P$.}
    \label{fig:clamping_cartoon}
\end{wrapfigure}

Energy clamping provides a formal framework by which dbCRN can perform inference. However, implementing energy clamping directly would involve carefully modifying the internal energies of different species in a dbCRN. Although theoretically possible, this is not an easy parameter to tune experimentally. In this section, we construct a dbCRN in a small volume $\nu$ can be coupled to a large external bath at volume $\mathbb{V} \gg \nu$ to produce a chemical system which implements energy clamping. We call these CRNs \textit{potentiated} dbCRNs. A physical illustration of a potentiated CRN can be found in figure \ref{fig:clamping_cartoon}. We will show that the species in the bath, denoted \textit{potential} species $\mathcal{P}$, produce a chemical potential equivalent to energy clamping.

Any dbCRN, $\mathcal{D}_G$, can be converted to a potentiated dbCRN, $\mathcal{D}^\mathcal{P}_G$, by adding an additional potential species $P_i$ that are coupled to a corresponding the clamped species $S_i \in \mathcal{S}^C$ and held at a constant concentration $[P_i]= \frac{p_i}{\mathbb{V}}$ where $p_i$ is a count. Formally, this can be viewed as replacing all instances of $S_i$ in each reaction (both inputs and outputs) with $S_i + P_i$.\footnote{Technically, the potential species $P_i$ could go on either side of the reaction - we consider just one side for simplicity.} For example, the dbCRN with reaction $2 S_i \xrightleftharpoons[]{} S_j + S_k$ become a potentiated dbCRN with reaction $2 S_i + 2 P_i \xrightleftharpoons[]{} S_j + S_k + P_j + P_k$ in the case that all species are clamped. Formally, given a dbCRN $\mathcal{D}_G = (\mathcal{S}, \mathcal{R}, k)$ we will denote the the potentiated dbCRN $\mathcal{D}^\mathcal{P}_G = (\mathcal{S}\cup\mathcal{P}, \mathcal{R}_\mathcal{P}, k)$ where $\mathcal{P}$ are the potential species, $\mathcal{R}_\mathcal{P}$ are the reactions from $\mathcal{R}$ modified to include potentials, and the rates $k_r$ of each reaction are unchanged. This model is a hybrid model in the sense that $S_i$ are measured in counts and vary stochastically but $P_i$ varies continuously and is measured in concentration. This formalism will be derived as the limiting case of a completely stochastic CRN when the count of $P_i$ becomes large in the SI. In other work, we couple the potentiated dbCRN back to the bath through a set of (non detailed balanced) chemical reactions producing a CRN which is able to autonomously learn from its environment. 

\textbf{Theorem:} Any potentiated dbCRN $\mathcal{D}^\mathcal{P}_G$ is detailed balanced. \textit{Proof:} See SI. Remark: it is not strictly necessary for each $S_i$ to have a unique potential species $P_i$; many $S_i$ could share the same $P_j$ (a universal potential species) if they are all clamped together simultaneously. Remark: if individual reactions are connected to a potential, instead of each species, the CRN is no longer detailed balanced and most of the results in this paper are not expected to hold. The physics of such systems have been studied elsewhere~\cite{schmiedl2007stochastic,polettini2014irreversible}.

\textbf{Theorem:} Let $\mathcal{D}^\mathcal{P}_G$ be a potentiated dbCRN with equilibrium distribution $\pi^P(s)$ and $\mathcal{D}_{G^C}$ be an energy clamped dbCRN with equilibrium distribution $\pi^C(s)$. $\pi^P(s) = \pi^C(s)$ provided that:
\begin{equation}
    \label{eq:potential_clamping_equivalence}
    \Delta_i = \mu_i = G^P_i + \log \frac{p_i}{\mathbb{V}} = G^P_i + \log [P_i]
\end{equation}
Here the energy difference $\Delta_i$ is equated to the chemical potential of $P_i$, commonly denoted $\mu_i$, and $G_i^P$ is the energy of $P_i$. \textit{Proof:} See SI \ref{SI:proof_of_potential_clamping_equivalence}.

\section{Thermodynamics of Inference via Clamping}
\label{sec:thermodynamics_of_clamping}

This section provides energetic and thermodynamic costs for inference with potentiated dbCRNs.  We start with a dbCRN $\mathcal{D}_G$ with equilibrium distribution $\pi_G$ and, via the clamping process, end up with a new distribution $\pi_{G'}$. The final distribution $\pi_{G'}$ can be physically realized in a variety of ways: by changing the energies $G' = G + \Delta$; or by equipping $\mathcal{D}_G$ with potential species so that $G' = G + \mu$. Importantly, in the both these scenarios, $\pi_{G'}$ remains a dbCRN. In the following section, we will investigate reversibly and non-reversibly modulating the potentials of a potentiated dbCRN.

To analyze the costs of inference, we consider clamping a potentiated dbCRN by changing the concentrations of the potential species directly by changing the external baths. In the following analysis, the \textit{system} will be a tuple $(\pi, P)$ where $\pi$ is an initial distribution of the CRN and $P$ are the potential bath concentrations. The system is always in thermal equilibrium with its environment (the solvent) at a temperature $T$. By changing the concentration of the external bath of the species $P$, the chemical potentials $\mu$ can be controlled. Similarly to the analysis by Ouldridge and collaborators\cite{ouldridge2018power,brittain2021would}, we will imagine a set of different baths $b$ each with a concentration of potential species $[P^b]$ which can be freely disconnected and reconnected reaction volume. The purpose of this analysis is to highlight different costs of inference in certain extreme conditions which can act as benchmarks for future investigations.
We begin with some basic thermodynamic definitions for a dbCRN at a (not necessarily equilibrium) distribution $\omega$ connected to a potential bath with concentrations $[P]$. Denote the energy function $\mathcal{G}^P$ which includes chemical potential terms. The internal energy,\footnote{For simplicity, we are assuming that the species energies $G_i$ are in fact enthalpies. In reality, they may contain both enthalpic and entropic components however for simplicity of presentation we ignore that complication.} $\mathbb{G}_\omega^P$, entropy, $\mathbb{S}_\omega^P$, and free energy, $\mathbb{F}_\omega^P$, of a dbCRN are given by:\cite{ouldridge2018importance}
\begin{align}
    \mathbb{G}_{\omega}^P = \langle \mathcal{G}^P(s) \rangle_{\omega} = \sum_s \omega(s) \mathcal{G}^P(s), \quad 
     \mathbb{S}_{\omega}^P = -k_B \langle \log \omega \rangle_{\omega} = -k_B \sum_s \omega(s) \log \omega(s), \quad \textrm{and} \quad 
     &\mathbb{F}_{\omega}^P = \mathbb{G}^P_{\omega} - T \mathbb{S}^P_{\omega}.
\end{align}
If $\omega = \pi^P$, the equilibrium distribution for potential concentrations $[P]$, then entropy and free energy can also be written:
\begin{align}
     \mathbb{S}^P_{\pi^P} = k_B (\frac{\mathbb{G}_{\pi^P}}{T} + \log Z_{\pi^P})
     \quad \quad
     \mathbb{F}^P_{\pi^P} = - k_B T \log Z_{\pi^P}
\end{align}
The free energy difference between a potentiated dbCRN with potential concentrations $[P]$ in a non-equilibrium distribution $\omega$ and its equilibrium $\pi^P$ is given by \cite{qian2001relative}:
\begin{align}
    \Delta \mathbb{F}^P_{\omega \to \pi^P} &= \mathbb{F}^P_{\omega} - \mathbb{F}^P_{\pi^P} = \mathbb{G}^P_\omega - T\mathbb{S}^P_\omega - \mathbb{F}_{\pi^P}
    = k_B T \sum_s \omega(s) (\log \omega(s) - \log \pi^P(s))
    = k_B T \mathbb{D}(\omega \mid \mid \pi^P).
    \label{eq:free_energy_is_dkl}
\end{align}
\begin{wrapfigure}{r}{.5\textwidth}
    \centering
    \includegraphics[width = .49\textwidth]{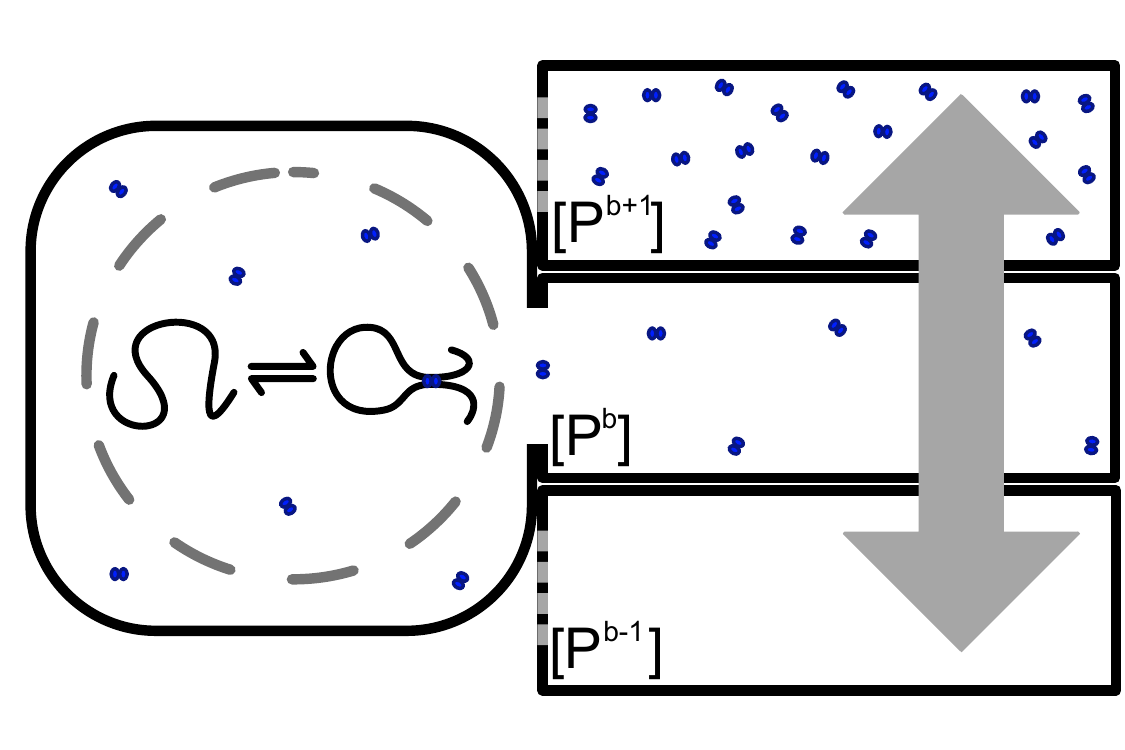}
    \caption{An illustration of inference via potential baths. The baths different concentrations of potential species (shown in blue) can be connected and disconnected from a to a dbCRN in a small volume (dashed circle) for free. Not to scale.}
    \label{fig:inference_cycle}
\end{wrapfigure}
\textbf{The Reversible Case:}
Following the analysis by Ouldridge and collaborators~\cite{ouldridge2018power}, we consider a process in which the potentiated dbCRN begins in an equilibrium distribution $\pi^{P^0}$ connected to a bath of potential species at concentration $[P^0]$. We now disconnect the potentiated dbCRN from this potential bath and instantaneously reconnect it to a new bath $[P^1]$ then allow the potentiated dbCRN to equilibrate to a new equilibrium distribution $\pi^{P^1}$. In our analysis we approximate the potential baths being constant; in reality, a small number of $P_i$ species may be transferred between potential baths which could result in additional entropy production terms due to mixing. This procedure is repeated a total of $N$ times resulting in a sequence of potential concentrations: $[P^0] \to ... \to [P^j] \to ... \to [P^N]$. Similarly, there is a corresponding sequence of equilibrium distributions: $\pi^{P^0} \to ... \to \pi^{P^j} \to ... \to \pi^{P^N}$. In each step, the non-equilibrium free energy difference given by equation (\ref{eq:free_energy_is_dkl}) will be dissipated into the bath \cite{ouldridge2018importance}:
\begin{equation}
    \label{eq:delta_F}
    \Delta \mathbb{F}_{\pi^{P^j} \to \pi^{P^{j+1}}}^{P^{j+1}} = k_B T \mathbb{D}(\pi^{P^j} \mid \mid \pi^{P^{j+1}}).
\end{equation}
Schematically a single step of the process can be described by:
\begin{equation}
    (P^j, \pi^{P^j}) \xRightarrow{\mathbb{W}} (P^{j+1}, \pi^{P^j}) \xRightarrow{\Delta \mathbb{F}} (P^{j+1}, \pi^{P^{j+1}})
\end{equation}
Notice that some work $\mathbb{W}$ must be done to modulate the potentials but we assume that is recoverable by reversing the process. Summing over all $N$ transitions results in the total entropy produced by the process. In the limit of $N \to \infty$, no entropy will be produced by this process: 
\begin{align}
    \Delta \mathbb{S}^{\textrm{reversible}} = \lim_{N\to \infty} \frac{1}{T} \sum_{j=0}^{N-1} \Delta \mathbb{F}_{\pi^{P^j} \to \pi^{P^{j+1}}}^{P^{j+1}} = 0.
\end{align}
\textit{Proof:} See SI \ref{SI:proof_of_thermodynamic_reversibility}. This shows that a quasi-static reversible process is capable of inference and generates no entropy. This result indicates that inference, if computed slowly, can be free---something we suspect might be advantageous to cellular life in certain conditions. 

\textbf{The Maximally Irreversible Case:} 
Next, we consider a non-reversible process. In this protocol, the potential baths are changed instantaneously from $P^0$ to $P^N$. This can be imagined as moving through the same series of baths as above but infinitely quickly instead of infinitely slowly (or equivalently moving directly to the final buffer).The CRN is initially at equilibrium with distribution $\pi^{P^0}$. Following the change, the CRN will initially be out of equilibrium at the distribution $\pi^{P^0}$ but at potential concentration $P^N$. Then, the CRN relaxes to the equilibrium distribution $\pi^{P^N}$ based upon the new potential species' concentrations. 
\begin{equation}
    (P^0, \pi^{P^0}) \xRightarrow{\mathbb{W}} (P^{N}, \pi^{P^0}) \xRightarrow{\Delta \mathbb{F}} (P^{N}, \pi^{P^{N}})
\end{equation}
Again, approximate the potential baths as constant and assume any work $\mathbb{W}$ to modulate the potentials is recoverable by the reverse process. Then, the entropy generated as the potentiated dbCRN equilibrates with the bath can be written by applying equation (\ref{eq:delta_F}) to a single step:
\begin{equation}
    \Delta \mathbb{F}_{\pi^{P^0} \to \pi^{P^{N}}}^{P^{N}} = k_B T \mathbb{D}(\pi^{P^0} \mid \mid \pi^{P^{N}}) \geq 0.
\end{equation}
In other words, the thermodynamic cost of performing inference \textit{quickly} is the relative entropy between the initial distribution of the system and the final distribution of the system. Importantly, performing inference in this way always has an energetic cost because the relative entropy is always positive. Entropy is produced by the dbCRN being pushed out of equilibrium and then settling into a new equilibrium state.

\section{Discussion}
We begin our discussion with the title of this paper---a metaphor between detailed balanced chemical reaction networks and Boltzmann machines. As a paradigm of generative machine learning models, BMs have three important features. First, they are capable of representing a broad class of distributions and make use marginalization over hidden hidden units to enhance their computational power. Second, BMs can condition these distributions to perform probabilistic inference via the clamping process. Finally, there is a simple algorithm which allows for their parameters to be trained from data. We argue that dbCRNs share these fundamental features. First, we analyzed necessary conditions for dbCRNs to represent complex distributions far from the canonical product Poisson form they are best known for and show that this can be accomplished by careful control of their species counts and initial conditions in order to restrict their reachability class. We then provided necessary conditions for dbCRNs to make use of hidden species to further increase the complexity of the distributions they generate. Addressing the second point, we showed how energy clamping is a form of conditioning and can enable complex inferential computations which make use of hidden units. However, in this work we have not yet addressed the third point---that a simple learning algorithm exists for dbCRNs---because previous work has already uncovered such an algorithm allowing for the species energies $G_i$ to be tuned \textit{in silico} \cite{poole2017chemical}:
\begin{equation}
    \label{eq:dbCRN_learning_rule}
    \frac{\textrm{d}G_i}{\textrm{d}t} = \frac{\textrm{d} \mathbb{D}(\overline{\pi} \mid \mid \pi)}{\textrm{d} G_i} = \epsilon (\langle s_i \rangle_{\overline{\pi}} - \langle s_i \rangle_{\pi}).
\end{equation}
Here the time derivative is meant to denote a gradient descent algorithm over training epochs, $\epsilon$ is the learning rate, $\pi$ is the distribution of a dbCRN, and $\overline{\pi}$ is a distribution of a dbCRN which has been clamped to a sample of the training data. In other work, we examine this learning algorithm in more detail and show how it can be implemented by non-detailed balanced CRNs.

We next turn to the question of the representational power of dbCRNs. We have provided a necessary condition for marginalization over hidden species to increase the complexity of distributions a dbCRN can represent: that the reachable states of the visible species must be dependent on the states of the hidden species. This leads to the question of which dbCRN architectures can produce far-from-Poisson distributions and how the representational power of dbCRNs relates to their underlying structure. So far, we know of two powerful constructions which require many species and/or reactions as well as carefully controlled binary species counts. The pixel CRN construction used to produce the face distribution in figure (\ref{fig:dist}) uses $O(N)$ hidden species where $N$ is the size of the supported visible distribution. Similarly, the chemical Boltzmann machine construction described in Section \ref{sec:cbms} uses $O(2^D K)$ reactions where $D$ is the degree of the underlying BM and $K = \log_2 N$ is the number of binary species needed to represent a distribution of size $N$. This leads us to speculate that restricting the reachability class of a dbCRN is a kind of computational resource much as entanglement is a resource in quantum computing~\cite{chitambar2019quantum}.

Next, we examined how modulating the energies of the species in a dbCRN can be interpreted as computing a conditional distribution and hence a form of inference. Energy clamping, as we named this process, can also be implemented by coupling clamped species in the dbCRN to potential baths.  Modulating the energies of chemical species can happen easily by changing the sequences of DNA or other polymers and therefore likely is a form of inference that can occur evolutionary. Using a chemical potential to modulate the mean of a chemical signal could also easily occur in a real or synthetic biochemical network and may be used for an organism to adapt to its environment. We suspect that these kinds of inferential systems could be realised \textit{in vivo} due to the binding and unbinding events between the genome and different proteins or RNAs such as occurs in epigenetic chromatin structures in eukaryotes and in folded or looping conformations caused by many transcription factors in bacteria and eukaryotes~\cite{becker2014bacterial,chiariello2016polymer}. In order to build such a system in the lab, the greatest challenge will be to produce a chemical network with low species counts and dynamically measure the stochastic fluctuations. Advances in droplet based technologies coupled with microfluidics~\cite{shang2017emerging} and positioning of single DNA molecules~\cite{gopinath2016engineering} provide potential avenues for the construction of such a circuit \textit{in vitro}. In a separate paper  we further examine how potential clamping can be implemented in a chemical environment. 

To conclude, we wish to emphasize that the mathematical structures described in this paper are in fact quite general, even though they may appear overly specific. At the highest level, detailed balanced CRNs are ubiquitous in biology and engineered bio-molecular systems; interactions such as molecular binding, diffusion, and conformation changes are frequently detailed balanced. Indeed, for our results on energy clamping and inference to hold, a detailed balanced sub-system need only be at a dynamic quasi-equilibrium relative to other subsystems. Similarly, our results on potentiated dbCRNs are also somewhat more general than they may appear: potentials may be shared between multiple species and, as exemplified in the DNA looping model, can occur on either side of a detailed balanced reaction. This flexibility means that dynamic and driven biochemical systems may, in fact, be generalized Boltzmann machines in disguise. Finally, the observation that dynamic changes in generalized potential species coupled to detailed balanced sub-systems could be acting as molecular machine learning models - representing complex distributions and computing inference while simultaneously adapting to and learning from the environment is addressed separately. 

\acknow{Acknowledgement: we would like to thank R.M. Murray of Caltech for helpful feedback on this project. W.P. was supported by the U.S. National Science Foundation. Additionally, the NSF in conjunction with the Indo-US Science and Technology Forum provided support for W.P. to spend a semester in Mumbai India working with M.G. which was instrumental to the success of this project.}

\matmethods{Methods: all numerical simulations were carried out using the Gillespie algorithm implemented in the Bioscrape python package \cite{Swaminathan121152}. Steady state distributions were found by simulating a CRN for a long time until the distribution converged.}

\showmatmethods{} 

\showacknow{} 


\bibliography{bib}

\section{Supplemental Information}
\subsection{Proof that Energy Clamping is Conditioning}
\label{SI:proof_of_conditional_limit}
The following is a proof of the theorem (\ref{eq:energy_clamping_conditional_limit}). \\

\textbf{Claim 1:} First, we will show that the distribution $P = \pi_C(s^F, s^C)$ minimizes the relative entropy $\mathbb{D}(P \mid \mid \pi)$ subject to the constraint that $\langle s^C \rangle_{P} = \overline{c}$ using the method of Lagrange multipliers. We first define the functional we wish to minimize:
\begin{equation}
    J(P) = \mathbb{D}(P \mid \mid \pi) +  \Delta \cdot \left(\overline{c} -  \sum_s   P(s)  s \right)+ \alpha \left( 1 - \sum_s P(s) \right).
\end{equation}
where $\Delta \in \mathbb{R}^N$ will constrain the means and $\alpha \in \mathbb{R}$ will normalize the distribution. Note that by setting $\Delta_i = 0$ the mean of $S_i$ will not be constrained, so this functional can constrain any subset of the species $S$. Optimize this function with respect to each component of the function space $P_s = P(s)$:
\begin{align}
    \frac{\partial J}{\partial P_s} 
    &= \frac{\partial}{\partial P_s} \left[ \left( \sum_{s'}  P_{s'} \log \frac{P_{s'}}{\pi_{s'}} + \Delta \cdot (\overline{c} - s') P_{s'} \right) + \alpha \left(1 - \sum_{s'}  P_{s'} \right) \right] &&= 0 \\
    &= \frac{\partial}{\partial P_s} [P_s \log \frac{P_s}{\pi(s)} + \Delta \cdot (\overline{c} - s) P_s + \alpha P_s] && = 0\\
    &= \log \frac{P_s}{\pi(s)} + 1 + \Delta \cdot (\overline{c} - s) + \alpha &&= 0 \\
    \implies P^*_s &= C \pi(s) e^{\Delta \cdot s} = \frac{1}{Z_C}e^{-\sum_i G_i^C s_i + \log s_i!} = \pi_C(s)
\end{align}
Where the $C$ is a normalizing constant which becomes part of the partition function $Z_C$ and $\Delta$ is chosen so that $\langle s^C \rangle_{\pi_C} = \overline{c}$. In practice, optimization techniques can be used to find $\Delta$ provided that $\overline{c} \in \overline{\Gamma}_{s^0}$ where $\overline{\Gamma}_{s^0}$ is the convex hull around $\Gamma_{s^0}$ because the mean does not have to be an integer value. The conceptual meaning of this result is that $\pi_C$ is optimal choice of distribution, with expected value $\langle s^C \rangle = \overline{c}$ for the clamped species $S^C$, to encode $\pi$.
\\\\
We are now ready to invoke the conditional limit theorem (See Cover and Thomas Elements of Information Theory theorem 11.6.2 p. 371\cite{cover1999elements}). This theorem states that if one were to observe a \textit{very rare} sample of $\pi$ with mean $\overline{c}$ the distribution of that sample would be given by $\pi_C$. Importantly, this provides a rigorous definition of what it means to condition upon the a distribution having a mean value which allows us to interpret energy clamping as a form of conditioning and hence inference. We note this theorem has been applied to understand similar chemical systems previously by Virinchi et al~\cite{virinchi2017stochastic}.
\\\\
\textbf{Define:} $\mathcal{P}$ to be the probability simplex (set of distributions) over the reachability class of the CRN $\Gamma_{s^0}$. 
\begin{equation}
    \mathcal{P} = \left\{ \mu: (s^F, s^C) \to \mathbb{R} \quad \textrm{s.t.} \sum_{s^F, s^C \in \Gamma_{s^0}} \mu(s^F, s^C) = 1\right\}.
\end{equation} Note that this is a convex subspace of a Banach function space~\cite{cover1999elements}.
\\\\
\textbf{Define:} The space of distributions in the probability simplex, $E \subseteq \mathcal{P}$, where the expected value of the species $s^C = \overline{c}.$
\begin{equation}
    E = \{\mu(s^F, s^C) \in \mathcal{P} \textrm{ s.t. }  \langle s^C \rangle_\mu = \overline{c}\}
\end{equation}
here $\langle \cdot \rangle_\mu$ denote the expected value relative to the distribution $\mu$.
\\\\
\textbf{Claim 2:} $E$ is convex. \textit{Proof:} the expected value $\langle \cdot \rangle_\mu$ is a linear operator so the constraint $\mathbb{E}_\mu[s^C] = \overline{c}$ defines an affine subspace of $P$. The intersection of an affine subspace and a convex space is also convex, so $E$ is convex~\cite{cover1999elements}.
\\\\
We now define a set of $N$ samples from the distribution $\pi$ and the empirical \textit{distribution of types} derived from the sample sequence.
\\
\textbf{Define:} $\mathbb{S}^n = s^1, ..., s^n$ are a vector of  $n$ i.i.d. samples from the distribution $\pi$D
\\\\
\textbf{Define:} $\mathbb{P}_{\mathbb{S}^n}$ is the empirical distribution of types derived from the sequence of samples $\mathbb{S}^n$.
\begin{equation}
    \mathbb{P}_{\mathbb{S}^n}(s) = \frac{\sum_i \mathbb{I}(s^i, s)}{n} 
\end{equation}
where $\mathbb{I}(s', s)$ denotes the identity operator.
\\\\
We can now apply the condition limit theorem which states that for any convex set $E$ and empirical distribution of types $\mathbb{P}_{\mathbb{S}^n}$ derived from a distribution $\pi$, in the in the limit $n \to \infty$ the probability of a sample $s^i$ having the value $s$ is given by:
\begin{align}
    \lim_{n \to \infty} \mathbb{P}(s^i = s \mid P_{X^n} \in E) = P^* \quad \quad P^* = \textrm{argmin}_{P \in E} \mathbb{D}(P \mid \mid \pi) = \pi_C
\end{align}
\\\\
In words, consider drawing a set of samples $n$ samples $\mathbb{S}^n$ from a detailed balanced CRN with equilibrium distribution $\pi$ where the means of a subset of the species $s^c$ of these samples are given by $\langle s^C \rangle_{\mathbb{P}_{\mathbb{S}^n}} = \overline{c}$. In the limit $n \to \infty$, the samples will be distributed according a distribution, $P^*(s^F, s^C)$, which can be produced by another detailed balanced CRN with equilibrium distribution $P^* = \pi_C$ where the species' energies $G_i^C = \Delta_i + G_i$ are chosen such that $\langle s^F \rangle_{\pi_C} = \overline{c}$. This process can be thought of as a kind of \textbf{sample clamping} because the sampling process must be somehow biased to produce a different mean. This technical description allows us to equate sample clamping with the energy clamping; they are equivalent descriptions of conditioning on the mean value of $\pi$:
\begin{align}
    \pi_C(s^F, s^C) = \pi(s^F, s^C \mid \langle s^C \rangle = \overline{c}).
\end{align}
And hence energy clamping produces a conditional distribution and therefore can be interpreted as a kind of inference.

\subsection{Proof of Energy Clamping and Potential Clamping Equivalence}
\label{SI:proof_of_potential_clamping_equivalence}
The following is a proof of the theorem (\ref{eq:potential_clamping_equivalence}):\\
\textbf{Claim 1} Any potentiated dbCRN is still detailed balanced with equilibrium distribution given by:
\begin{equation}
\label{eq:potentiated_dbCRN_full}
    \pi^P(s, p) = \frac{1}{Z} \exp -(\sum_i G_i s_i + G^P_i p_i + \log s_i! + \log p_i!)
\end{equation}
\textit{Proof:} Consider the reaction $\sum_i I_i S_i \xrightleftharpoons[k^-]{k^+} \sum_i O_i S_i$. After being connected to the potential baths, this reaction becomes:
\begin{align}
    &\sum_i I_i (S_i + P_i) \xrightleftharpoons[k^-]{k^+} \sum_i O_i (S_i + P_i)\\
    &\frac{k^+}{k^-} = e^{-\sum_i (G_i + G^P_i)(O_i-I_i)} \\
    &\rho^+(x) = k^+ \prod_i \frac{s_i! p_i!}{(s_i - I_i)!(p_i - I_i)!}
    \quad \quad 
    \rho^-(x) = k^- \prod_i \frac{s_i! p_i!}{(s_i - O_i)!(p_i - O_i)!}
\end{align}
$I$ and $O$ are the reactions inputs and outputs which are the same stoichiometry for each species $S_i$ and its potential $P_i$ by construction. $G_i$ and $G^P_i$ are the energies of the species $S_i$ and $P_i$, respectively. $\rho^{\pm}(x)$ are the propensities of the forward and backward reactions. Note that the rate constants (in units of per second) are unchanged by the addition of the potential species by construction. Next, we show that $\pi^P(s+O-I) \rho^-(s+O-I) = \pi^P(s) \rho^+(s)$ proving that $\pi^P$ is an stationary distribution which satisfies detailed balance:
\begin{align}
    &\frac{\rho^+(s)}{\rho^-(s+O-I)}
    = \frac{k^+}{k^-}\frac{\prod_i \frac{s_i! p_i!}{(s_i - I_i)!(p_i - I_i)!}}{\prod_i \frac{(s_i+O_i-I_i)! (p_i+O_i-I_i)!}{(s_i - I_i)! (p_i - I_i)!}}
    = \prod_i \frac{e^{(G_i + G_i^P)(I_i - O_i)} s_i! p_i !}{(s_i + O_i - I_i)! (p_i + O_i - I_i)!} \\
    &\frac{\pi^P(s+O-I, p+O-I)}{\pi^P(s)} 
    = \prod_i \frac{s_i! p_i! e^{-G_i(s_i + O_i - I_i) - G_i^P (p_i + O_i - I_i)}}{(s_i + O_i - I_i)! (p_i + O_i - I_i)! e^{-G_i s_i - G_i^P p_i}} = \prod_i \frac{e^{(G_i + G_i^P)(I_i - O_i)} s_i! p_i !}{(s_i + O_i - I_i)! (p_i + O_i - I_i)!}.
\end{align}
Because the dbCRN with potential species is detailed balanced, we can simply apply the product Poisson formula (\ref{eq:product_poisson}) to get the equilibrium distribution (\ref{eq:potentiated_dbCRN_full}). 
\\\\
\textbf{Claim 2:} $\pi^P(s, p)$ is just a function of $s$.
\begin{equation}
\label{eq:potentiated_dbCRN_s}
    \pi^P(s) = \frac{1}{Z} e^{-\sum_i G_i s_i - G^P_i (s_i - s_i^0 + p_i^0) - \log s_i! - \log (s_i-s_i^0+p_i^0)!}
\end{equation}
\textit{Proof:} Given an initial condition $(s^0, p^0)$, it is clear that the change in $s_i$ and $p_i$ are coupled by the construction of the CRN: $p_i = s_i - s_i^0 + p_i^0$. Equation (\ref{eq:potentiated_dbCRN_s}) results from inserting this conservation law into (\ref{eq:potentiated_dbCRN_full}). Furthermore, note that the terms $G_i(s_i - s_i^0 + p_i^0) - log(s_i - s_i^0 + p_i^0)!$ are a kind of stochastic chemical potential.
\\\\
\textbf{Claim 3:} In the limit $p_i \gg 0$ and $p_i \gg s_i - s_i^0$, $\pi^P$ has the simplified form:
\begin{equation}
    \label{eq:potentiated_dbCRN_approx}
    \pi^P(s) \approx \frac{1}{Z} e^{-\sum_i G_i s_i - G^P_i s_i - \log s_i! - s_i \log p_i^0}
\end{equation}
\textit{Proof:} Using Stirling's approximation, $\log(s_i - s_i^0 + p_i^0)! \approx (s_i - s_i^0 + p_i^0) \log(s_i - s_i^0 + p_i^0)$. If $p_i^0 \gg s_i - s_i^0$ then $\log(s_i - s_i^0 + p_i^0) \approx \log p_i^0$. Finally, the constant terms $G_i^P(p_i^0 - s_i^0)$ and $(p_i^0 - s_i^0) \log p_i^0$ will factor out between the Gibbs factor and the partition function.
\\\\
\textbf{Claim 4:} Equating the exponential terms of (\ref{eq:potentiated_dbCRN_approx}) to the exponential terms of an energy clamped CRN (\ref{eq:energy_clamp_equilibrium}) term by term results in the relation:
\begin{equation}
    G_i^C = G_i + G_i^P + \log p_i^0 = G_i + \mu_i.
\end{equation}
where $\mu_i$ is the chemical potential of $P_i$ and is retrieved by changing from units of counts to units of concentration. \textit{Proof:} this follows easily from some simple algebraic manipulation.

\subsection{Proof of the Thermodynamically Reversible Limit}
\label{SI:proof_of_thermodynamic_revesibility}
In this section, we will prove that the thermodynamically reversible limit of an infinitely slow clamping process exists and generates no entropy. We will consider a dbCRN $\mathcal{D}_G$ and slowly change the energies of the CRN to $G+\Delta$. We note that modulating the energies $\Delta$ is equivalent to modulating the potentials $P$ and results in simpler notation. We will assume $\Delta$ changes over $N$ steps resulting in a sequence of equilibrium distributions for entire process: $\pi^0,...,\pi^j,...,\pi^N$. Each distribution is given by:
\begin{align}
    \pi^j(s) = \frac{1}{Z^j}e^{-(G + \frac{j \Delta}{N})\cdot s - \log s!} \quad \textrm{and} \quad Z^j = \sum_s e^{- (G + \frac{j \Delta}{N})\cdot s - \log s!}. 
\end{align}
Here, $\cdot$ denotes the dot product between two vectors and the factorial of a vector is the product of the factorialization of its components: $s! = \prod_i s_i!$.
First we will apply equation \ref{eq:free_energy_is_dkl} iteratively to each step of the $N$ steps in the process:
\begin{align}
    \frac{\Delta \mathbb{S}_N}{k_B}
    &= \sum_{j=0}^{N-1} \mathbb{D}(\pi^{j} \mid\mid \pi^{j+1})
    =\sum_{j=0}^{N-1} \sum_s \frac{1}{Z^j}e^{-(G + \frac{j \Delta_i}{N}) \cdot s - \log s!} \log \frac{Z^{j+1}}{Z^j} \frac{e^{-(G + \frac{j \Delta}{N}) \cdot s - \log s!}}{e^{-(G + \frac{(j+1) \Delta}{N}) \cdot s - \log s_i!}} \\
    &= \sum_{j=0}^{N-1} \sum_s \pi^j(s) (\log \frac{Z^{j+1}}{Z^j} + \log e^{\frac{\Delta}{N} \cdot s})
    = \sum_{j=0}^{N-1} \sum_s \pi^j(s) (\log \frac{Z^{j+1}}{Z^j} + (\frac{\Delta}{N} \cdot s)) \\
    &= \sum_{j=0}^{N-1} \log \frac{Z^{j+1}}{Z^j} + \langle \frac{\Delta}{N} \cdot s \rangle_{\pi^j}
\end{align}
Next we simplify the ratio of partition functions, $\frac{Z^{j+1}}{Z^j}$:
\begin{align}
    \frac{Z^{j+1}}{Z^j} 
    &= \frac{\sum_s e^{- (G + \frac{(j+1) \Delta}{N}) \cdot s - \log s!}}{\sum_s e^{- (G + \frac{j \Delta}{N}) \cdot s - \log s!}}
    = \sum_s \frac{e^{- (G + \frac{j \Delta}{N}) \cdot s - \log s!}e^{- \frac{\Delta}{N} \cdot s}}{Z^j} 
    = \sum_s \pi^j(s) e^{- \frac{\Delta}{N} \cdot s} = \langle e^{- \frac{\Delta}{N} \cdot s} \rangle_{\pi^j} \\
\end{align}
Combining these results we obtain the total entropy produced for an $N$-step process:
\begin{align}
    \frac{\Delta \mathbb{S}_N}{k_B}
    = \sum_{j=0}^{N-1} \left[ \log \langle e^{-\frac{\Delta}{N} \cdot s}\rangle_{\pi^j} + \langle \frac{\Delta}{N} \cdot s \rangle_{\pi^j} \right].
\end{align}
Next, we take the limit $N \to \infty$. For the quasi-static reversible process:
\begin{align}
    \Delta \mathbb{S}^{\textrm{reversible}} = \lim_{N \to \infty} \Delta \mathbb{S}_N
\end{align}
We begin by finding upper bounds for the both term in the sum:
\begin{align}
    A = \max_{s} e^{-\Delta \cdot s} \quad \textrm{and} \quad B = \max_s \Delta \cdot s
\end{align}
This allows the terms inside the sum to be bounded:
\begin{align}
    \log \langle e^{-\frac{\Delta}{N} \cdot s}\rangle_{\pi^j} \leq  \log \langle A^{\frac{1}{N}} \rangle_{\pi^j} = \frac{1}{N} \log A
    \quad \textrm{and} \quad
    \langle \frac{\Delta}{N}\cdot s \rangle_{\pi^j} \leq \langle \frac{B}{N} \rangle_{\pi^j} = \frac{B}{N}.
\end{align}
Furthermore, the infinite sums of the bounds converge:
\begin{align}
    \lim_{N\to \infty} \sum_{j=0}^{N-1} \frac{1}{N} \log A = \log A
    \quad \textrm{and} \quad
    \lim_{N \to \infty} \sum_{j=0}^{N-1}\frac{B}{N} = B
\end{align}
These results allow us to apply the dominated convergence theorem \cite{schechter1996handbook} which shows that the limit $N \to \infty$ and the summation commute. We will then apply the monotone convergence theorem \cite{schechter1996handbook} to pull the limit inside of the expected value for the first term; because $e^{-\frac{\Delta}{N}\cdot s}$ is monotone for positive $N$, the limit can be pulled inside the expected value. Similarly, for the second term $\frac{1}{N}$ is monotone in $N$ so the limit can be pulled inside the expected value:
\begin{align}
    \lim_{N\to \infty} \sum_{j=0}^{N-1}  \log \langle e^{-\frac{\Delta}{N} \cdot s}\rangle_{\pi^j}
    &= \sum_{j=0}^{N-1} \lim_{N\to \infty} \log \langle e^{-\frac{\Delta}{N} \cdot s}\rangle_{\pi^j}
    = \sum_{j=0}^{N-1}  \log \langle \lim_{N \to \infty}e^{-\frac{\Delta}{N} \cdot s}\rangle_{\pi^j}
    = 0.
    \\
    \lim_{N \to \infty} \sum_j^{N-1} \langle \frac{\Delta}{N}\cdot s\rangle_{\pi^j} 
    &= \sum_j^{N-1} \lim_{N \to \infty} \langle \frac{\Delta}{N}\cdot s\rangle_{\pi^j} 
    = \sum_j^{N-1} \langle \lim_{N \to \infty} \frac{\Delta}{N}\cdot s\rangle_{\pi^j} = 0.
\end{align}
This completes the proof that $\lim_{N \to \infty} \mathbb{S}_N = 0$, showing that the quasi-reversible limit exists for energy-clamping arbitrary detailed balanced CRNs.
\end{document}